\documentstyle[12pt,fleqn]{article}
\begin{document} 

\vspace*{2cm}
\begin{center}
{\Large 
Randomly Amplified Discrete Langevin Systems 
}
\end{center}


\begin{center}
{
Nobuko Fuchikami \par
{\it Department of Physics, Tokyo Metropolitan University \par
Hachioji, Tokyo, 192-0397, Japan} \par
{\small Email: fuchi$@$phys.metro-u.ac.jp} \par
\vspace*{1cm}
(February 2, 1999) 
}
\end{center}



\vspace*{1cm}

\noindent
{\bf Abstract} \par

\vspace{2mm}

A discrete stochastic process involving random amplification with additive 
noise is studied analytically. 
If the non-negative random amplification factor $b$ is such that
$<b^{\beta}>=1$ where $\beta$ is any positive non-integer,
then the steady state probability density function for the process 
will have power law tails of the form
$p(x) \sim 1/x^{\beta +1}$.  
This is a generalization of recent results for $0 < \beta < 2$ obtained by 
Takayasu et al. in Phys. Rev. lett. {\bf 79}, 966 (1997).
It is shown that 
the power spectrum of the time series $x$ becomes Lorentzian,   
even when $1 < \beta < 2$, i.e., in case of divergent variance.

\vspace*{1cm}


\newpage

\setlength{\baselineskip}{9mm}

Power law behavior of distribution function is widely observed 
in nature \cite{bak97}.
Recently, Takayasu et al. presented a new general mechanism leading to 
the power law distribution \cite{tst97}.
They analyzed a 
discrete stochastic process which involves random amplification together with 
additive external noise.
They clarified necessary and sufficient conditions to 
realize steady power law fluctuation with divergent variance using 
a discrete version of linear Langevin equation expressed as
%
\begin{equation}
x(t+1)=b(t)x(t)+f(t),
\label{model}
\end{equation}
where $f(t)$ represents a random additive noise 
and $b(t)$ is a non-negative stochastic coefficient.
They derived the following time evolution equation for
the characteristic function $Z(\rho,t)$ which is the Fourier 
transform of the probability
density $p(x,t)$:
%
\begin{equation}
Z(\rho, t+1)= \int_{0}^{\infty} W(b) Z(b \rho, t) db \, \Phi (\rho) ,
\label{z}
\end{equation}
where $W(b)$ is the probability density of $b(t)$ and $\Phi (\rho)$ is
the characteristic function for $f(t)$.
They showed that when $\left<b^{\beta}\right >=1$ holds for $0< \beta <2 $, 
the 
second moment 
$\left< x^2(t)\right>$ diverges as $t \rightarrow \infty$, but
Eq. (\ref{z}) has
a unique steady and stable solution :    
\begin{equation}
\lim_{t \rightarrow \infty} Z(\rho,t) \equiv Z(\rho)=
1- {\mbox {const}}\times |\rho|^{\beta} + \cdots, 
\end{equation}
which yields 
the power law tails in the steady probability density
\begin{equation}
\lim_{t \rightarrow \infty} p(x,t) \equiv 
p(x) \sim 1/x^{\beta +1},
\end{equation}
or equivalently, the cumulative distribution
\begin{equation}
P(\ge|x|) \sim 1/x^{\beta}.
\end{equation}

They also made numerical simulations of Eq. (\ref{model}) by 
employing a discrete
exponential
distribution for $W(b)$, and showed that the theoretical estimate of the
relation between $\beta$ and the parameters 
specifying $W(b)$ (Eq. (15) in \cite{tst97}) nicely
fits with the simulation ``even out of the range of 
applicability, $\beta > 2$''. 
They state that ``The reason for this lucky coincidence is not clear'',
although they point out at the same time
that the power law distribution tails are a generic property
of Eq. (\ref{model}) \cite{tst97}. 

In this Brief Report, the following two statements will be presented:

\vspace*{3mm}
(A) Takayasu et al.'s theory can be straightforwardly extended for $\beta>2$: 
If $\left<b^{\beta}\right>=1$ holds for a positive 
non-integer $\beta$, then
there exists a unique steady and stable solution of Eq.(\ref{z})
\begin{equation} 
Z(\rho) = \sum_{m=0}^{n}A_{2m}(-1)^{2m} \rho^{2m}
-C|\rho|^{\beta}+O(\rho^{2n+2}), 
\label{zgeneral}
\end{equation}
where $2n$ is the largest even number that is smaller than $\beta$.
This $Z(\rho)$ leads to $p(x) \sim 1/x^{\beta+1}$.

\vspace*{3mm}
(B) When $\left<b^{\beta}\right >=1$ for 
a non-integer $\beta$ between 1 and 2,
the power spectral density (PSD) of $x(t)$ is Lorentzian increasing with the 
observation time $T$ as
\begin{equation}
\ S(\omega,T)  \sim \frac{2}{T} 
\frac{x_0^2 }{\ln \left<b^2 \right>} 
\frac{(1/\tau_1) \left<b^2\right>^T}{(1/\tau_1)^2 + \omega^2} 
\quad \quad {\rm for}\quad T \gg 1,
\label{spectrum}
\end{equation}
where 
\begin{equation}
x_0^2 \equiv \left<x^2(0) \right> + 
\frac{\left<f^2 \right>}{\left<b^2 \right>-1},
\end{equation}
\begin{equation}
\tau_1 = \frac{1} {\ln \left<b^2\right>+ \ln[1/\left<b\right>]}.
\end{equation}

From the statement (A), 
``the coincidence'' 
found in \cite{tst97} is naturally understandable.
To prove (A), 
we assume the following form for $Z(\rho)$:
\begin{equation}
Z(\rho)=\sum_{n=0}^{\infty} a_n \rho^{n}
+ |\rho|^{\beta} \sum_{n=0}^{\infty} c_n \rho^n, 
\quad \quad  a_0 \equiv 1,
\label{zgeneral1}
\end{equation}
and substitute it into Eq. (\ref{z}) in the 
limit $t \rightarrow \infty$.  
If $\Phi (\rho)$ is an even function (i.e., the distribution function 
of $f(t)$ is symmetric as assumed in \cite{tst97}), we can first prove that
$a_1 =0$ because $\left<b\right> \ne 1$. Also, $c_1=0$ because
$\left<b^{\beta +1}\right> \ne 1$. 
Thanks to $a_{2 m-1}=0$ and $\left<b^{2 m+1}\right> \ne 1$, 
$a_{2 m+1}=0$ is derived. 
Similarly, 
$c_{2 m-1}=0$ and $\left<b^{\beta + 2 m+1}\right> \ne 1$ yield
$c_{2 m+1}=0$.   
We can thus prove that $a_n$ and $c_n$ in Eq. (\ref{zgeneral1}) vanish
for all odd numbers $n$, i.e., Eq. (\ref{zgeneral}) holds.
(Note that the $n$ th moment $\left< x^n(t)\right>$ with $n>\beta$ diverges
not only for even number $n$ but also for odd number which corresponds to 
the vanishing coefficient $a_n$.)
Taking exactly the same procedures as in \cite{tst97}, 
we can prove that this solution is
unique and stable.
In case of $\beta > 2$,
we have a finite variance but higher order
moments, $\left< x^n(t)\right>$ with $n > \beta$, diverge as
$t \rightarrow \infty$.

To derive the probability density $p(x)$, we
only need to assume that all $k$-th derivatives of 
$Z(\rho)$ satisfy the boundary condition 
\begin{equation}
\lim_{\rho \rightarrow \pm \infty} d^{k}Z(\rho)/d \rho^{k} =0.
\label{bc}
\end{equation}
Using Eq. (\ref{bc}), we can partially integrate the expression
\begin{equation}
p(x) \equiv \frac{1}{2 \pi}
\int_{- \infty}^{\infty} e^{ix \rho} Z(\rho) d \rho 
\end{equation}
$[\beta]+1$ times, where $[\beta]$ is the 
largest integer that is smaller than $\beta$. 
Thus we obtain the asymptotic 
expansion as
\begin{equation}
p(x) \sim |x|^{-(\beta +1)} \int_{- \infty}^{\infty} e^{-i \xi}
|\xi|^{\beta - [\beta] -1} d \xi
\sim |x|^{-(\beta +1)} \Gamma(\beta -[\beta]),
\end{equation}
where $\Gamma$ is the Gamma function. 

To prove the statement (B), we note that the two-time correlation function
is rigorously obtained from Eq. (\ref{model}):
\begin{equation}
\phi (\tau,t) \equiv \left<x(t+ \tau) x(t)\right> = \left<x^2(t)\right>
\left<b\right>^{\tau},
\end{equation}
where
\begin{equation}
\left<x^2(t)\right> =
\left<b^2\right>^t \left<x^2(0)\right> + 
\frac{1-\left<b^2\right>^t}{1-\left<b^2\right>} \left<f^2\right>.
\end{equation}
%

If $1<\beta <2$, we have a relation 
$0< \left < b \right> < 1 < \left<b^2 \right>$
because the function $G(\gamma) \equiv \left<b^{\gamma}\right>$ satisfies
$G(0)=1$ and 
$G''(\gamma)>0$ \cite{tst97}.
Then $\phi$ increases with $t$, but decays with
$\tau$ as 
$\sim e^{-\tau/\tau_0}$ 
for any fixed value of $t$ (Debye-type relaxation),
with the relaxation time
\begin{equation}
\tau_0 = \frac{1} {\ln [1/\left<b \right>]}.
\end{equation}
Since the correlation function depends on both $\tau$ and $t$, 
the Wiener-Khinchin relation cannot be used to obtain the PSD. 
Defining the PSD which depends on the 
observation time $T$ as
\begin{eqnarray}
S(\omega,T) & \equiv & 
\left<\left| \int_0^{T} e^{i \omega t} x(t) dt \right| ^2\right>/T
\nonumber
\\
& = & 2{\rm Re} \left\{\int_0^{T}d \tau \int_0^{T- \tau} dt
e^{i \omega \tau} \left<x(t+\tau) x(t)\right>\right\}/T,
\end{eqnarray}
and using $\phi(\tau, t)$ obtained above, we arrive at the expression 
(\ref{spectrum}). The spectrum is $1/f^2$ -type for $f \gg 1/\tau_1$ and 
flat for $f \ll 1/\tau_1$.
Equation (\ref{spectrum}) implies that the power increases exponentially 
with the observation time $T$, which 
corresponds to
the divergent behavior of the variance $\left<x^2(t)\right>$.
(We have neglected the case $0 < \beta < 1$ where even the average of 
$x$ diverges as 
$\left<x(t) \right> = \left<b \right>^t \left<x(0) \right>$ 
because $\left<b \right>>1$.)

When $\beta>2$, both $0< \left<b\right> < 1$ and 
$0< \left<b^2\right> < 1$ hold and results are rather trivial:
\begin{eqnarray}
\left<x^2\right> & \equiv & \lim_{t \rightarrow \infty} \left<x^2(t)\right> =
\frac{1}{1-\left<b^2\right>} \left<f^2\right>,\\
\phi(\tau) & \equiv & \lim_{t \rightarrow \infty} \phi(\tau, t) =
\left<x^2\right> \left<b\right>^{\tau},\\
S(\omega) & \equiv & \lim_{T \rightarrow \infty}S(\omega, T)
= 2 \left< x^2 \right> \frac{(1/\tau_0)}{(1/\tau_0)^2+\omega^2}.
\end{eqnarray}
Thus, as far as the PSD is measured, we cannot observe any singular
aspect, higher order singularities being hidden.


The stochastic process described by Eq. (1) generally leads to 
the power law behavior 
$p(x) \sim 1/x^{\beta+1}$, while it also yields a Lorentzian spectrum
$S(\omega) \propto 1/ [ (1/\tau)^2 + \omega^2 ]$.
A colored noise, or $1/f^{\alpha}$ fluctuation, whose PSD is 
proportional to $1/\omega^{\alpha}$ has attracted much attention since
$1/f$ noise was discovered several decades ago\cite{johnson25}.
Such power law behavior of the PSD is also observed widely in nature, and
these two power laws, one in the 
probability density and the other in the PSD, are sometimes discussed 
together[2].
Therefore it is interesting to know whether an extremely long time scale $\tau$
can be involved in the present stochastic process. 
Because, in that case, 
the observation time $T$, which relates with the low
frequency cut-off $\omega_0$ as $\omega_0 = 2\pi/T$, cannot reach this 
time scale, then a $1/f^2$ fluctuation, namely
$S(\omega) \sim 1/\omega^2$ (for $\omega \ge \omega_0$) is 
observed $\it practically$.

One can immediately see that the time constant $\tau_0$ or $\tau_1$
becomes large in very limited cases.
First, the average of $b$ should be close to unity, i.e., 
$\left<b\right> =1-\epsilon$ with $0<\epsilon \ll1$. Then 
$\tau_0$ becomes $\sim 1/\epsilon \gg 1$. 
Furthermore, in case of $\beta>2$, we need $\left< b^2 \right>$ smaller 
than unity, while 
in case of $1<\beta<2$, the condition 
$\left<b^2 \right> =1+\delta$ with $0<\delta \ll1$ is necessary.
In the latter case, we obtain $\tau_1 \sim 1/(\epsilon + \delta) \gg 1$.
The exponential or Poisson distribution for $W(b)$ 
does not lead to such a long time constant. 
One example of large $\tau_1$ is obtained by choosing
$W(b)$ 
to be a narrowly peaked distribution
having the average which is slightly smaller than unity and the second moment
slightly larger than unity.

As pointed out in the above, it should be noted that a stochastic process
whose stationary density function has power law tails will not necessarily 
exhibit the power law behavior 
in PSD.


\end{document}